\documentclass[review]{elsarticle}
\usepackage{amsmath,amssymb,amsfonts,amsthm}
\newcommand{\BR}{\mathbb{R}}
\newcommand{\be}{\begin{equation}}
\newcommand{\ee}{\end{equation}}
\usepackage{graphicx}
\begin{document}
\begin{frontmatter}
\title{What is Time in Quantum Mechanics?}
\author[myaddress]{Arkadiusz Jadczyk}
\address[myaddress]{Quantum Future Group Inc.\fnref{qfgaddress}}
\fntext[qfgaddress]{2955 Rte de Toulouse, 82100 Castelsarrasin, France}
\ead{kairos@quantumfuture.net}
\begin{abstract}
Time of arrival in quantum mechanics is discussed in two versions: the classical axiomatic ``time of arrival operator" introduced by J. Kijowski and the EEQT method. It is suggested that for free particles the two methods may
lead to the same result. On the other hand the EEQT method can be easily geometrized within the framework of Galilei-Newton general relativistic quantum mechanics developed by M. Modugno and collaborators, and it can be applied to non-free evolutions. The way of geometrization of irreversible quantum dynamics based on dissipative Liouville equation is suggested.
\end{abstract}
\end{frontmatter}
\section{Introduction: Why ``time''}
In standard quantum mechanics time is a parameter in Schr\"odinger's equation for the wave functions. Wave functions there should be square integrable over space. We rarely integrate over time. So, there is no canonical ``time operator" in quantum mechanics, while we do have position, momenta, and energy operators.
There is an evident asymmetry between space and time in quantum mechanics. Certain asymmetry is also present in classical mechanics and field theory. The fundamental equations are hyperbolic, the initial conditions (Cauchy's data) are data ``at a given time''. But in quantum mechanics time is also related to specifically quantum--mechanical problem of ``measurement''. Measurements are usually considered as ``instantaneous in time''. We are measuring physical quantities at different times, and we are interested in ``time evolution'' of these quantities. Yet the question of ``how to measure time in quantum mechanics?'' is asked by physicists and philosophers again and again.
\cite{muga021,muga092}
\subsection{But which time?}
J. P. Dowling, discusses quantum time measurements in his recent monograph reviewing modern quantum technologies \cite{dowling2013} --- therefore ``quantum time'' becomes important not only as an object interest for mathematical physicists. Dowling also speculates that \cite{dowlingblog}
\begin{quotation}
 ... there is some ur-theory, likely a phenomenological one, which unifies
non-relativistic quantum theory and non-quantum relativity theory. (...) some intermediate unified theory between quantum gra\-vity and what
we have now and that this theory in certain limits produces non-relativistic quantum theory and
non-quantum relati\-vity theory.\\
\end{quotation}
Diosi and Lukacs \cite{diosi1987,diosi1989} suggested the need to create a unified theory of Newtonian Quantum Mechanics and Gravity.

An elegant, pure geometrical, formulation of Newton--Galilei general relativistic quantum mechanics was pioneered by Marco Modugno (with the participation of the present author) in 1993 \cite{jadczyk-modugno1993a,jadczyk-modugno1993b}.
\subsection{Geometry of Galilei-Newton relativity\label{sec:gnst}}
Space--time, in this formulation,  is a refined version of that of Galilei and
of Newton, i.e. space--time with absolute simultaneity. In particular,
four dimensional space--time manifold $E$ of \textit{events} is fibrated over one--dimensional
time $B.$ The fibers $E_t$ of $E$ are three--dimensional
Riemannian manifolds, while the basis $B$ is an affine space over
$\BR.$ Coordinate systems $x^\mu=(x^0,x^i) ,$ $i=1,2,3\, ,$ on $E$
are adapted to the fibration. In adapted coordinates any two events with the same coordinate $x^0$ are
simultaneous, i.e. they are in the same fibre of $E .$

Coordinate transformations between any two adapted coordinate systems
are of the form:
\begin{eqnarray}
x^{0'}&=&x^0+const ,\nonumber \\
x^{i'}&=&x^{i'}\left(x^0,x^{i}\right) .
\label{eq:adapted}\end{eqnarray}
Let $\beta$ be the \textit{time form}:
$$\beta = dx^0 .$$

In adapted coordinates we have $\beta_0=1,\beta_i=0$.
$E$ is equipped with a \textit{contravariant degenerate metric tensor} which,
in adapted coordinates, takes the form
\be \begin{pmatrix}
0&0&0&0\\
0&g^{11}&g^{12}&g^{13}\\
0&g^{21}&g^{22}&g^{23}\\
0&g^{31}&g^{32}&g^{33}
\end{pmatrix},\ee
where $g^{ij},\, (i,j=1,2,3)$ is of signature $(+++).$ We denote by $g_{ij}$ the
inverse $3\times 3$ matrix. It defines Riemannian metric on the three--dimensional
fibers of $E .$

Let us consider torsion--free affine connections $\Gamma$ in $E,$ together with the associated covariant derivative $\nabla,$ that preserves
$g^{\mu\nu}$ and $\beta$:
\be (\nabla g)^{\mu\nu}=0,\label{eq:ngmn}\ee
\be (\nabla \beta)_\mu = 0.\label{eq:nbm}\ee
 What is the freedom in choosing such a connection?

The condition (\ref{eq:nbm}) is equivalent to the conditions
\be \Gamma^0_{\mu\nu}=0\ee on the connection coefficients. Let us
introduce the notation \be \Gamma_{\mu\nu,i}=g_{ij}\Gamma^j_{\mu\nu}.\ee
Then the condition  (\ref{eq:ngmn}) is equivalent to the equations:
\be
\partial_\mu g_{ij}=\Gamma_{\mu i,j}+\Gamma_{\mu j,i} .
\ee
Now, because of the assumed zero torsion, the space part of the
connection can be expressed in terms of the three-dimensional space metric in
the Levi-Civita form:
\be
\Gamma_{ij,k}=\frac{1}{2}\left( \partial_i g_{jk}+
\partial_j g_{ik}-
\partial_k g_{ij}\right) .
\ee
From the remaining equations:
\be
\partial_0 g_{ij}= \Gamma_{0i,j} + \Gamma_{0j,i}
\ee
we find that the symmetric part of $\Gamma_{0i,j}$ is equal to
$\frac{1}{2}\partial_0 g_{ij} ,$ otherwise the connection is
undetermined. We can write it as
\be
\Gamma_{i0,j}=\frac{1}{2}\left(\partial_0 g_{ij}+\Phi_{ij}\right) ,
\ee
\be
\Gamma_{00,j}=\Phi_{0j} ,
\ee
where $\Phi_{\mu\nu}=-\Phi_{\mu\nu}$ is an arbitrary \textit{antisymmetric
object}.
It is then natural to introduce quantities $\mathbf{E},\mathbf{B}$ defined by
\be E_i=\Phi_{0i},\, B_i=\epsilon_{ijk}\Phi_k,\, (i=1,2,3).\ee
Assuming that the fibers of the space--time manifold $E$ are flat, that is in
some adapted coordinates we have $g_{ij}=\delta_{ij},$ and performing special Galilei transformation:
\begin{eqnarray}
\mathbf{x}'&=&\mathbf{x}-\mathbf{v}t\\
t'&=& t,
\end{eqnarray}
we easily find that
\begin{eqnarray} \mathbf{E}'&=&\mathbf{E}+\mathbf{v}\times\mathbf{B}\\
\mathbf{B}'&=&\mathbf{B}.
\label{eq:eb}\end{eqnarray}
There are now two ways of interpreting these degrees of freedom in the connection. First we may notice that the transformation laws (\ref{eq:eb}) are the same as in the ``electric limit" of Galilean electromagnetism \cite{bellac1973,rousseaux2013}. Therefore it is tempting to interpret $\mathbf{E}$ and $\mathbf{B}$ as proportional to the electric and magnetic fields in Galilean electrodynamics. But such an interpretation would force us to choose different connections for particles with different ratios of $e/m.$ There is however a different interpretation: $\mathbf{E}$ and $\mathbf{B}$ belong to the universal force of gravitation in \textit{gravitoelectromagnetism}, as it is discussed, for instance, in \cite{dematos2006,dematos2007}. This second interpretation seems to be more natural.

Let $J_1E$ be the affine jet bundle $J_1E\stackrel{\pi}{\longrightarrow}E.$
\footnote{Jets at $x\in E$ can be, in this case, identified with tangent vectors $y^\mu=(y^0,y^i)$ at $x$ for which $y^0=1.$} We can parametrize $J_1E$ by coordinates $(x^\mu,y^i).$ $J_1E$ carries the canonical form $\theta$ given by
\be \theta^i=dx^i -y^i \,dx^0 .\ee

The connection $\Gamma$ induces, in a natural way, an affine connection in   $J_1E,$ therefore it defines a one--form $\nu_\Gamma$ on $J_1E$ with values in the vector bundle $VE$ of vectors tangent to the fibers of $E.$ We can define then the two--form $\Omega$ on $J_1E:$
\be
\Omega=g_{mn}\,\nu_\Gamma^m\wedge\theta^n .\ee
One can show that the form $\Omega$ is closed,  $d\Omega=0,$ if and only if the curvature tensor $R$ of $\Gamma$ satisfies additional requirements:
\be R^{\mu \phantom{\nu}
\sigma}_{\phantom{\mu}  \nu \phantom{\sigma}  \rho} =
R^{\sigma\phantom{\rho}  \mu}_{ \phantom{\sigma} \rho\phantom{\mu} \nu},\ee
where
\be R^{\mu\ \phantom{\nu}\sigma}_{\phantom{\mu}
 \nu \phantom{\sigma} \rho}
 =g^{\mu\lambda}
R_{\lambda \nu \phantom{\sigma} \rho}^{\phantom{\mu \nu}\sigma} .\ee
This happens to be equivalent to the condition on $\Phi$ of being closed:
\be \partial_{[ \mu} \Phi_{\nu\sigma]}=0.\label{eq:dphi}\ee
It can be verified by a direct calculation that the condition (\ref{eq:dphi}) is covariant with respect to the transformations (\ref{eq:adapted}) between adapted frames, even though $\Phi_{\mu\nu}$ is not a tensor.
\subsection{Quantization}
With space--time geometry encoded as above, quantization procedure is straightforward. The arena for quantization is a principal $U(1)$ bundle $Q$ over $E$ and its pullback $Q^\dagger$ to $J_1E.$ Among principal connections on $Q^\dagger$ there is a special class of connections, namely those whose connection forms vanish on vectors tangent to the fibers of $Q^\dagger\rightarrow Q.$ Quantization is accomplished by selecting a connection $\omega$ in this class for which the curvature form is $i\Omega.$ In coordinates such a connection is of the form:
\be \omega = i\left( d \phi + a_{\mu} dx^{\mu}\right),\ee
where $0\leq \phi \leq 2\pi$ parametrizes the fibres of $Q$,
$$
\begin{array}{l}
a_{0} = - \frac{1}{2} g_{ij}y^i y^j + A_{0} , \cr\cr
a_{i} =  g_{ij} y^{j} + A_{i} ,
\end{array}
$$
and $A_{\nu} = \left(A_{0},A_{i}\right)$ a local
potential for $\Phi:$
\be \Phi_{\mu\nu}=\partial_\mu\,A_\nu - \partial_\nu\,A_\mu.\ee
Schr\"odinger's equation can then be interpreted in terms of the parallel transport (over time) with respect to the induced connection in the bundle of Hilbert spaces over the fibers of $E.$ Details and extensions can be found in the comprehensive review \cite{janyska2001} and references therein.
\subsection{Time of events}
The above geometrical formulation of quantization is well adapted for describing the continuous evolution in time of wave functions and expectation values of physical observables. But already in 1913, that is long before quantum theory as we know it today was invented,  Niels Bohr suggested that
there are discontinuous transitions between stationary states of electrons in atoms - in other words: \textit{quantum jumps}. While we can't see electrons jumping from one orbit to another, we can register photons emitted as a result of these jumps. These registration acts are \textit{events}, and we can record their time. Therefore in quantum theory \textit{events}, together with their timing, are important observational data. Events are being recorded also in nuclear decays. Yet timing of the events was escaping precise quantum mechanical formulation, mainly because in the mathematical formalism of quantum mechanics \textit{time is a parameter, not an operator}. There are good reasons for this: we never measure \textit{time}, we measure \textit{time of events}. But in order to do it, we need to specify first what kind of events we are looking at. They should be \textit{physical events} of some kind, not just abstract mathematical points of space--time continuum.

In 1972 Eugene P. Wigner addressed this problem in his paper ``On the Time-Energy Uncertainty Relation" \cite{wigner72}. There he introduced the concept of \textit{time of arrival at a state}. However Wigner did not solve the problem, and, after careful examination, we can easily notice mathematical and logical errors in his expressions.\footnote{But, quoting from Irving John Good, a British brilliant mathematician, who worked as a cryptologist with Alan Turing: ``\textit{It is often better to be stimulating and wrong than boring and right.}'' \cite[p. 1]{good1962}}.

In 1974 two papers appeared addressing the problem of measuring time of events in quantum theory.

One possible solution to this annoying problem was proposed by V. S. Olkhovsky, E. Recami and A. J. Gerasimchuk in their 1974 paper ``Time Operator in Quantum Mechanics'' \cite{olkhovsky1974}, where the authors wrote:
\begin{quotation}
``\textit{... The fact that the operator {\font\larm = larm1000\larm\char 190}time{\font\larm = larm1000\larm\char 191} seems to have  peculiar (even if not exceptional) features($\,{}^*$) led to its unjustified neglect. As a consequence, the Heisenberg uncertainty correlations for energy and time got particular obscurity as compared to other ones.\\
($\,{}^*$) We shall see that it does {\em not} admit a spectral decomposition, in nonrelativistic quantum mechanics ...}''
\end{quotation}
While this approach, via Hermitian but non--selfadjoint operators, is still being actively pursued (see e.g. the review article \cite{olkhovsky09}), it is not the approach I will elaborate upon in the present paper.
\subsection{Kijowski appears in time}
In the same year, 1974, another classical paper on the subject of time in quantum mechanics was published by J. Kijowski \cite{kijowski74}. Let us demonstrate the essence of Kijowski's time operator on a simple toy model: free Schr\"odinger's particle in one space dimension.  Using atomic units in which mass of the particle $m=1$ and Planck's constant $\hbar=1$ Schr\"odinger's equation reads:
\be \Psi\in L^2(\BR),\, i\partial \Psi_t/\partial t=H\Psi_t,\ee
with
\be (H\Psi)(x)=-\,\frac{1}{2}\,\frac{\partial^2\Psi(x)}{\partial x^2}.\ee
Then $H=H^*,$ and the equation has a formal solution
\be \Psi_t=e^{iHt}\Psi_0,\, ||\Psi_t||=const.\ee
Kijowski considered the event of \textit{particle crossing the point $x=0$}, and proposed a solution that he also proved to be a unique one under a number of natural geometrical conditions. Kijowski's solution goes as follows.

Let $\tilde{\psi}(k)$ be the Fourier transform of $\Psi_0(x):$
\be \tilde{\psi}(k)=\frac{1}{\sqrt{2\pi}}\int_{-\infty}^\infty \Psi_0(x)\,e^{-ikx}\,dx.\ee
Define:
\be \psi^+(\tau)=\frac{1}{\sqrt{2\pi}}\int_{0}^\infty \sqrt{k}\,\tilde{\psi}(k)\,e^{\frac{-ik^2\tau}{2}}\,dk,\ee
\be \psi^-(\tau)=\frac{1}{\sqrt{2\pi}}\int_{-\infty}^0 \sqrt{-k}\,\tilde{\psi}(k)\,e^{\frac{ik^2\tau}{2}}\,dk,\ee
Then the probability of the event of crossing $x=0$ at time $\tau$ is given by the formula:
\be p(\tau)=|\psi^+(\tau)|^2+|\psi^-(\tau)|^2.\ee
The two terms in the above formula correspond to particles arriving at $x=0$ from the left and from the right respectively.
\subsubsection{Example: free Gaussian packet\label{sec:kij}}
\begin{figure}[ht!]
\centering
      \includegraphics[width=\textwidth]{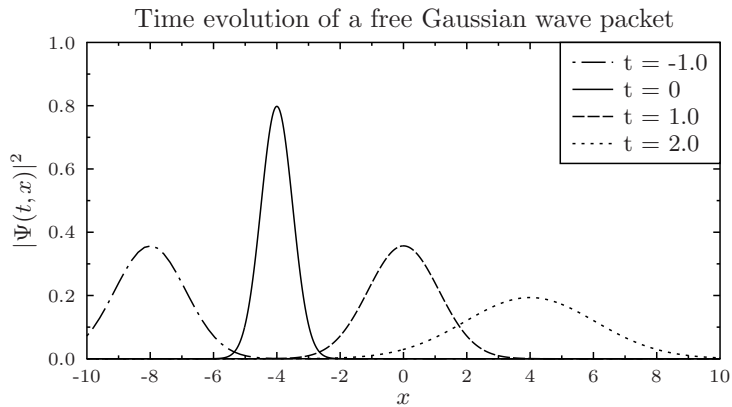}
  \caption{Free motion of the Gaussian wave packet $\Psi(t,x).$}\label{fig:gauss1}
\end{figure}

\begin{figure}[ht!]
\centering
      \includegraphics[width=\textwidth]{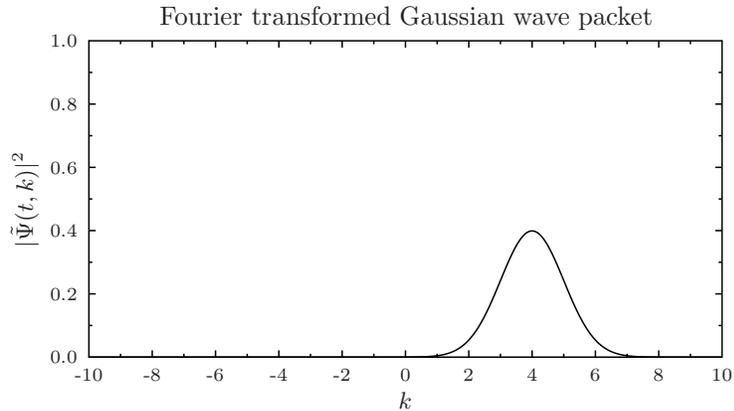}
  \caption{Fourier transformed evolution $\tilde{\Psi}(t,k)$}\label{fig:gauss2}
\end{figure}

Consider the following Gaussian wave packet
\be\Psi(0,x)=\sqrt[4]{\frac{2}{\pi }}\, e^{-(x+4)^2+4 i x+16 i}\ee
It is centered at $x=-4$ and its center moves with velocity $v=4$ to the right. We can write the solution of the free Schr\"odinger's equation with this initial condition explicitly:
\be\Psi(t,x)=\frac{\sqrt[4]{\frac{2}{\pi }} \exp \left(\frac{-8 t+ i (x+4)^2+4 (x+4)}{2 t- i}\right)}{\sqrt{1+2 i t}}.\ee
The center of this wave packet
moves at time $t=1$ to the origin $x=0.$ (Fig. \ref{fig:gauss1}).
Its Fourier transform $\tilde{\Psi}$ defined by
\be \tilde{\Psi}(t,k)=\frac{1}{\sqrt{2\pi}}\int_{-\infty}^\infty\,\Psi(t,x)e^{-ikx}\,dx\ee
keeps its shape constant in time. Only its phase (not shown in Fig. \ref{fig:gauss2}) oscillates.
For a Gaussian wave packet Kijowski's amplitudes $\psi^+(\tau)$ and $\psi^-(\tau)$ can be computed explicitly in terms of Bessel functions. However these explicit expressions are rather complicated and do not give us any insight into their behavior. It is better to represent them graphically.
\begin{figure}[ht!]
\centering
      \includegraphics[width=\textwidth]{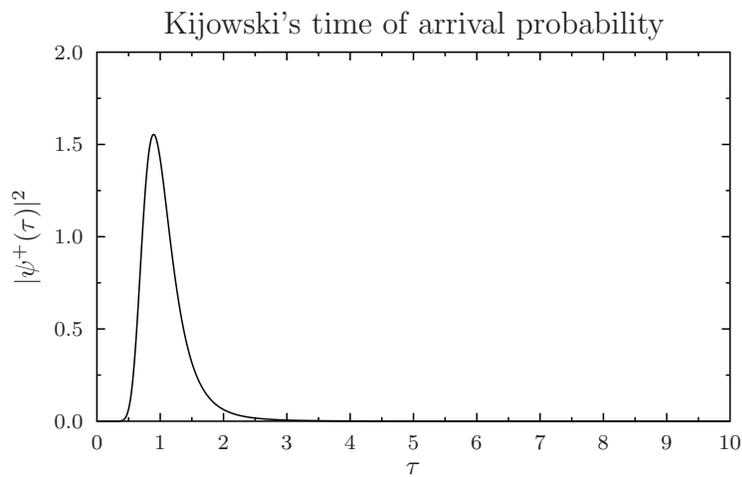}
  \caption{Right mover time of arrival distribution}\label{fig:phip}
\end{figure}
\begin{figure}[ht!]
\centering
      \includegraphics[width=\textwidth]{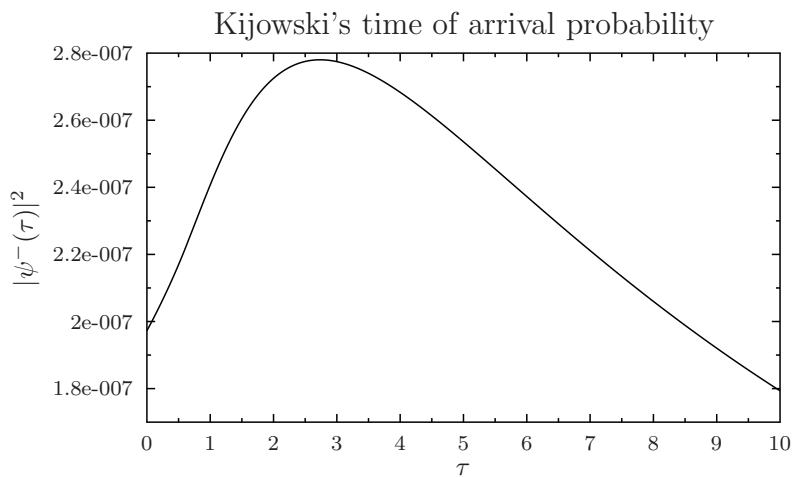}
  \caption{Left mover time of arrival distribution}\label{fig:phim}
\end{figure}
From Figs. \ref{fig:phip}, \ref{fig:phim} one can see that $\psi^+(\tau)$ behaves in an expected way: it has its maximum around $\tau=1.$ Indeed, it would take $\tau=1$ for a classical particle with velocity $v=4$ to move from $x=-4$ to $x=0.$ The amplitude of the probability distribution from $\psi^-(\tau)$ is so small that it can be neglected (it mainly comes from the part of the Gaussian distribution that is on the right of $x=0$ and has negative momentum component).
\subsubsection{Critics and replies}
Kijowski proved that his ``Time of Arrival" is unique under certain
well defined mathematical conditions. His solution, though generally accepted as mathematically sound, was criticised on other grounds. Grot, Tate and Rovelli \cite{grot1996} criticised Kijowski's solution in these words:
\begin{quotation}
``\textit{Kijowski \cite{kijowski74} obtained a probability distribution, but not on the usual Hilbert space; thus the interpretation of the wave function in terms of familiar quantities is
obscure.}''
\end{quotation}
Delgado and Muga \cite{delgado1997} repeated much the same:
\begin{quotation}
``\textit{Our results turn out to be similar to those previously
obtained by Kijowski \cite{kijowski74}. However, the approach by Kijowski was based on the definition of a nonconventional
wave function ...}''
\end{quotation}
Kijowski countered in \cite{kijowski1999}:
\begin{quotation}
``\textit{... I want to stress that the classification ‘‘nonconventional
wave function...whose relation to the conventional wave
function is unclear’’ could only be conceived by somebody
who did not read my paper carefully ...}''
\end{quotation}
More serious objections came from Bogdan Mielnik \cite{mielnik2005}\footnote{In 1994 Mielnik stated and analyzed a more general ``Screen Problem'' in Quantum Mechanics \cite{mielnik94}.}, who summarized the situation as follows:
\begin{quotation}
``\textit{It thus seems, that the axioms about the time of arrival omit quite
a number of physical aspects. It brings little comfort that they give
a unique probability. On the contrary, it brings new difficulties.}''
\end{quotation}
Kijowski, in reply \cite{kijowski2005}, essentially agreed with Mielnik:
\begin{quotation}
``\textit{... My construction of ``arrival time'' is indeed {\bf mathematically unique and final} within the conceptual framework of the standard interpretation of Quantum Mechanics. But I always considered it as an argument for further analysis of the conceptual framework of quantum theory. ...\\
Unfortunately at the moment there is no measurement theory, which could replace this (naive and very unsatisfactory!) picture. I wish Bogdan Mielnik to find one.}''
\end{quotation}
Apart from the seriously motivated objection raised by Mielnik, there is also another issue here, related to the subject of this paper: Kijowski's ``time of arrival'' heavily depends on the fact that we are dealing with free propagation in flat space and does not seem to be directly applicable in the presence of external potentials --- c.f. \cite[p. 10]{seidel2005} and references therein. Moreover it essentially depends on Fourier transform, and Fourier transforms do not translate easily from flat spaces to curved manifolds. Therefore it is rather improbable that Kijowski's time of arrival can be adjusted to a geometrical framework of quantum mechanics in general Galilei-Newton space-times outlined in
section \ref{sec:gnst}.\\
If so, what other options do we have?
\section{Event Enhanced Quantum Theory (EEQT): Time of Events}
True ``geometrical quantization'' must join two branches of mathematics: geometry and probability. While geometrical part is well developed, the probabilistic part is, till now, mostly neglected. Quantum theory is a theory of measurements, and measurements are irreversible processes that do not
necessarily destroy objects. Quantum mechanics, therefore, must include irreversibility. Quoting from Ilya Prigogine \cite{prigogine1997}:
\begin{quotation}
``\textit{I believe that we are at an important turning point in the history of science. We have come to the end of the road paved by Galileo and Newton, which presented us with an image of time--reversible, deterministic universe. We now see the erosion of determinism and the emergence of a new formulation of laws of Physics.}''
\end{quotation}
In this section I will propose a way of including irreversibility  and measurements into a geometrical formulation of quantum mechanics in a Galilei-Newton space--time. My suggestion is based on ``Event Enhanced Quantum Theory'' (EEQT) described, for instance, in \cite{blaja95a}.
\subsection{Main concepts of EEQT}
EEQT preserves a general algebraic scheme of quantum mechanics (Hilbert spaces, algebras of operators, states), but without its a priori physical interpretation. Physical interpretation follows there from dynamics. Dynamics is irreversible. It can be described mathematically at two different (but equivalent) levels. Either probabilistically, on the level of single systems, or, statistically,  on the level of ensembles of systems. For single systems
the Schr\"odinger equation is modified if measurements are taking place.
We have stochastic \textit{quantum jumps} separating periods of a continuous evolution. Jumps are accompanied by changes of pointer positions on measuring devices. This description requires the machinery of stochastic processes and it does not constitute an easy entry point for geometrization.

The alternative description, on the statistical ensemble level, along the ideas championed by Ilya Prigogine, requires, I believe, only adding to the present repertoire of geometrical tools, a few other tools that have already been developed in differential geometry, although for a different reason.
\subsection{Time of arrival according to EEQT\label{sec:leeqt}}
\begin{figure}[ht!]
\centering
      \includegraphics[width=\textwidth]{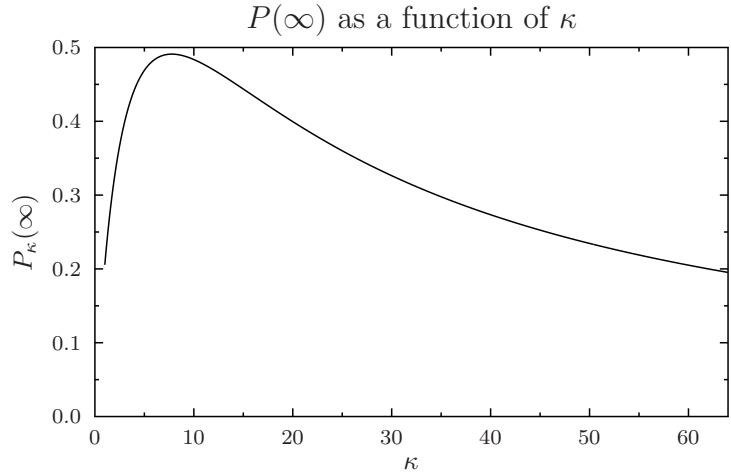}
  \caption{Total probability of detection.}\label{fig:pinf}
\end{figure}

\begin{figure}[ht!]
\centering
      \includegraphics[width=\textwidth]{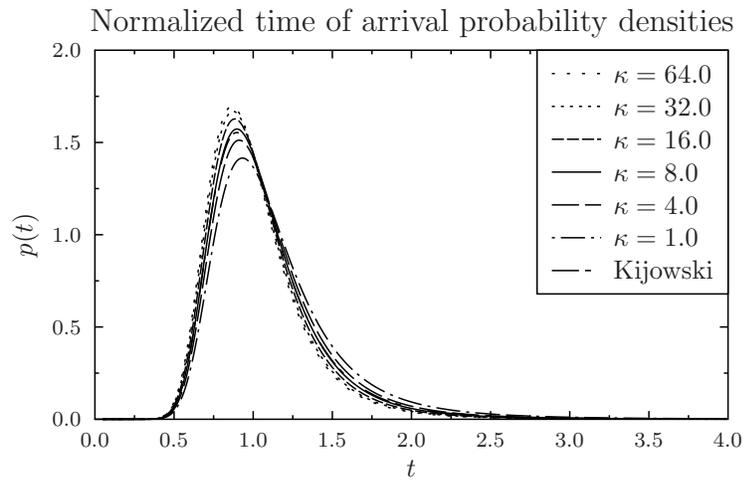}
  \caption{Normalized arrival times for different $\kappa$.}\label{fig:arnor}
\end{figure}

In EEQT a detector is characterized by a sensitivity parameter $\kappa>0.$ Here let us compare time of arrival obtained form EEQT with that of Kijowski. With the same configuration as in section \ref{sec:kij}, and with the idealized Dirac's delta detector at $x=0,$ using the formulas from Ref. \cite{blaja95f} (cf. also \cite{blaja97ar,palao}), we obtain (numerically) probability distributions shown in Fig. \ref{fig:arun}:
\begin{figure}[ht!]
\centering
      \includegraphics[width=\textwidth]{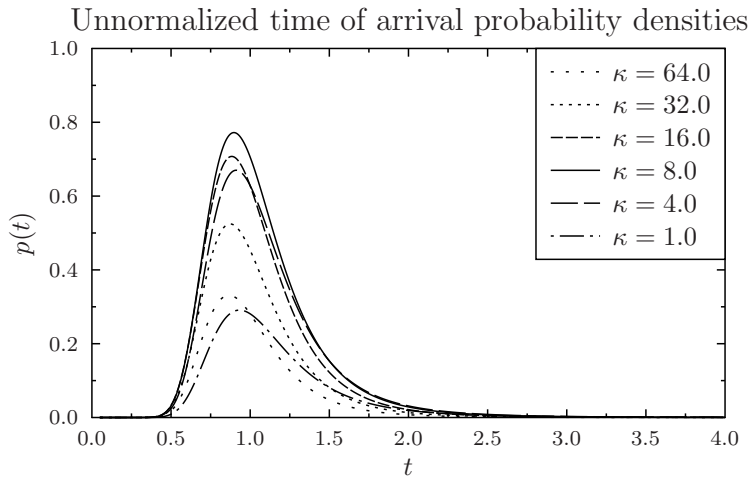}
  \caption{Time of arrival according to EEQT.}\label{fig:arun}
\end{figure}
These are unnormalized probabilities - the probability $P(\infty)$ that the particle will be detected in finite time is smaller than one. Some particles (wave packets)  will pass the screen without being detected, some will be reflected without triggering the detector. The value of $P(\infty)$ depends on the sensitivity parameter $\kappa,$ as can be seen in Fig. \ref{fig:pinf}.
It is then natural to normalize the probability curves - they will then represent the probability curves of those particle only that trigger a detection event. The normalized probability densities are show in Fig. \ref{fig:arnor}.
It can be seen form Fig. \ref{fig:pinf} that there is an optimal value of $\kappa$ for which $P(\infty)\approx 0.5.$ This value, for our Dirac delta detector, happens to be (numerically) twice the velocity of the Gaussian wave packet, in our case $\kappa=8.0.$
\begin{figure}[ht!]
\centering
      \includegraphics[width=\textwidth]{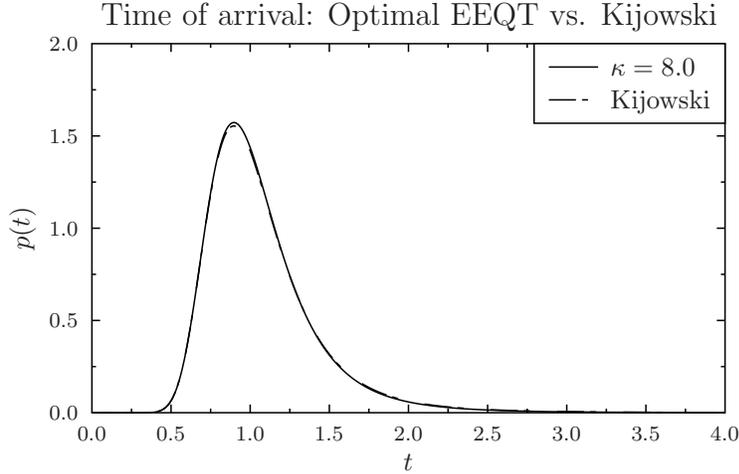}
  \caption{Is the difference between the two curves only due to the numerical approximation? }\label{fig:cmp}
\end{figure}
Comparing now the optimal EEQT time of arrival normalized probability curve with that of Kijowski's we can see that they almost indistinguishable. Perhaps they are exactly the same, and the small difference shown in Fig. \ref{fig:cmp} is the results of numerical approximation? This question needs further research.
\subsection{Geometrization of the Liouville equation?}
Quantum theory is a statistical theory, therefore certain elements of probabilistic machinery are necessary whenever models are to be compared with experiment. Usually this is done via Born's interpretation of quantum probability amplitudes, but Born's interpretation is an additional axiom that does not follow from the dynamics. Also, if we want to take into account measurement processes, additional problems appear. Ilya Prigogine advocated what he called a ``Unified Formulation of Quantum Theory" that would take into account, from the very beginning, the inherent irreversibility of event creation which is the basis of any observation, in particular observation of time of arrival.

Following Prigogine's ideas the fundamental mathematical object is the ``density matrix'' and the fundamental differential equation is the Liouville equation. Schr\"odinger's equation does not describe quantum jumps, one has to use a separate stochastic mechanism for that. But Liouville's equation can take into account the presence of measuring devices. It is a differential equation, and it should be possible to give a geometric meaning for Liouville's equation in a general Galilei-Newton geometrical background. I will now provide few ideas about how this can be done.

\subsubsection{Liouville's equation}
In the standard flat space formulation of quantum mechanics a particle detector is described by an operator $F,$ which can explicitly depend on time $t.$ In the simplest case $F$ is an operator of multiplication by a non-negative function of space point:
\be (F_t\Psi)(x)= f_t(x)\Psi(x).\ee
Quantum mechanical statistical state is described by a ``density matrix'' (or ``mixed state'') $\rho_t.$ $\rho,$ at each time $t,$ is a positive operator of trace one. The relation between wave functions and density matrices is such that to each wave function (quantum state) we can associate a density matrix - the orthogonal projection operator onto this state. Such density matrices describe pure states. In general, however, a density matrix does not correspond to a pure state. Without any measurements, when the dynamics is reversible and described by a self--adjoint Hamilton operator $H$ (for simplicity let us assume that $H$ does not depend explicitly on time), Schr\"odinger's equation can be equivalently written in terms of the time-dependent density matrix as follows:
\be \frac{d\rho_t}{dt}=-i[H,\rho_t].\label{eq:liou1}\ee
Eq. (\ref{eq:liou1}) is known as the Liouville form of the quantum mechanical state evolution. One can easily check that such an evolution preserves the purity of states. It is completely equivalent to the Schr\"odinger equation except for one fact: quantum mechanical effects such as, for instance, Aharanov-Bohm effect, or even simple double slit experiment, are harder to ``explain'' in the density matrix formalism, where the phase of the wave function is not explicitly represented. Feynman's method of superposition of amplitudes leads to the results much easier.

When there are measuring devices around, quantum dynamics becomes irreversible. Time evolution is no longer given in the form (\ref{eq:liou1}), pure states evolve, in general, into mixed states. For the case of one detector described by an operator $F_t$ (not necessarily Hermitian) the Liouville equation has additional terms. It takes the form
\be \frac{d\rho}{dt}=-i[H,\rho]+F_t^\dagger\rho F_t-\frac{1}{2}\{F_t^\dagger F_t,\rho\},\label{eq:liou2}\ee
 where the curly bracket stands for the anticommutator.
 One can easily check that this equation preserves both positivity and trace of $\rho.$
 It is this form of the Liouville equation that I propose as a good candidate for geometrization.\footnote{In the case of the Dirac delta counter located at $x=0,$ discussed in section \ref{sec:leeqt}, $F$ is an ``improper" operator of ``multiplication" by $\sqrt{\kappa}\delta(x)$ and $f^2$ is the ``multiplication" by $\kappa\delta(x).$ Of course, as it stands, it does not make sense mathematically, but it does make sense with a proper approach (limiting procedure) - the results are finite, as a physicist would expect.}
 \subsubsection{Geometrization of density matrices}
The Hamiltonian operator and the detector operator are both local, therefore they can be rather easily expressed in terms of local geometrical objects. It is not so with a general density density matrix. Assuming however that fibers of the Galilei--Newton space--time $E$ are compact Riemannian manifolds, we can assume that $\rho$ at any given time $t$ is an integral operator defined by a kernel function $\rho(x,y):$
\be (\rho\Psi)(x)=\int_{E_t} \rho(x,y)\Psi(y)dV(y),\ee
where $dV$ is the volume form of the Riemannian metric on the fibre. In such a form it should be now possible to express the dissipative quantum mechanics encoded in Eq. (\ref{eq:liou2}) in purely geometrical terms.

Of course we will have to deal now with two--point geometrical objects, but the path here was marked out long ago. A. Einstein and V. Bargmann discussed two--point tensor fields in Ref. \cite{einstein1944a,einstein1944b}, while J. L. Synge \cite[Ch. 2]{synge1950} derived many important properties of the two--point ``world function" in his formulation of General Relativity Theory.
\section{Conclusions}
Paraphrasing J. Kijowski {\it ``Unfortunately at the moment there is no measurement theory, which could replace this (naive and very unsatisfactory!) picture. I wish Bogdan Mielnik to find one."} --- I would rather say:
\begin{quotation}Fortunately at the moment there are measurement theories which could replace this naive and very unsatisfactory (orthodox) picture. I wish more mathematicians and mathematical physicists would get involved in this research. \\Geometry is pretty. Probability, on the other hand, is exciting, and it shows the way towards even prettier (conformal) geometry, and more satisfactory physics.\end{quotation}
\section{Acknowledgments}
I would like to thank Daniel Canarutto for the invitation and for his kind hospitality during the Workshop ``Geometry and Quantum Theories".

\end{document}